\documentclass[reprint,aps,prl,superscriptaddress,showpacs,footinbib,longbibliography,amsmath]{revtex4-1}
\pdfoutput=1
\usepackage{graphicx}
\usepackage{bm}
\usepackage{amssymb}

\usepackage{color}
\usepackage[colorlinks,bookmarks=false,citecolor=darkblue,linkcolor=red,urlcolor=blue]{hyperref}
\def\cred{
}
\definecolor{darkred}{rgb}{0.7,0.0,0.0}

\definecolor{darkblue}{rgb}{0,0.02,0.45}

\definecolor{darkgreen}{rgb}{0.02,0.45,0.0}

\definecolor{violet}{rgb}{0.8,0.2,0.6}

\newcommand{\Ha}{$H\,\|\,a$}
\newcommand{\Hb}{$H\,\|\,b$}
\newcommand{\Hc}{$H\,\|\,c$}
\newcommand{\Hpb}{$H\perp b$}

\begin{document}

\title{Anisotropic temperature-field phase diagram of single crystalline $\beta$-Li$_2$IrO$_3$: magnetization, specific heat, and $^7$Li NMR study}

\author{M. Majumder}
\email{mayukh.cu@gmail.com}
\author{F. Freund}
\author{T. Dey}
\affiliation{Experimental Physics VI, Center for Electronic Correlations and Magnetism, University of Augsburg, 86159 Augsburg, Germany}

\author{M. Prinz-Zwick}
\author{N. B{\"u}ttgen}
\affiliation{Experimental Physics V, Center for Electronic Correlations and Magnetism, University of Augsburg, 86159 Augsburg, Germany}

\author{Y. Skourski}
\affiliation{Dresden High Magnetic Field Laboratory (HLD-EMFL), Helmholtz-Zentrum Dresden-Rossendorf, 01328 Dresden, Germany}

\author{A. Jesche}
\author{A. A. Tsirlin}
\email{altsirlin@gmail.com}
\author{P. Gegenwart}
\affiliation{Experimental Physics VI, Center for Electronic Correlations and Magnetism, University of Augsburg, 86159 Augsburg, Germany}


\begin{abstract}
Detailed magnetization, specific heat, and $^7$Li nuclear magnetic resonance (NMR) measurements on single crystals of the hyperhoneycomb Kitaev magnet $\beta$-Li$_2$IrO$_3$ are reported. At high temperatures, {\cred anisotropy of the magnetization is reflected by the different Curie-Weiss temperatures for different field directions}, in agreement with the combination of a ferromagnetic Kitaev interaction ($K$) and a negative off-diagonal anisotropy ($\Gamma$) as two leading terms in the spin Hamiltonian. At low temperatures, magnetic fields applied along $a$ or $c$ have only a weak effect on the system and reduce the N\'eel temperature from 38\,K at 0\,T to about 35.5\,K at 14\,T, with no field-induced transitions observed up to 58\,T on a powder sample. In contrast, the field applied along $b$ causes a drastic reduction in the $T_N$ that vanishes around $H_c=2.8$\,T giving way to a crossover toward a quantum paramagnetic state. $^7$Li NMR measurements in this field-induced state reveal a gradual line broadening and a continuous evolution of the line shift with temperature, suggesting the development of local magnetic fields. The spin-lattice relaxation rate shows a peak around the crossover temperature 40\,K and follows power-law behavior below this temperature. 
\end{abstract}

\maketitle

\section{Introduction} 
Strong spin-orbit coupling in compounds of $4d$ and $5d$ transition metals triggers large exchange anisotropy that gives rise to {\cred magnetic scenarios uncommon to $3d$ compounds, where Heisenberg or Ising exchanges usually prevail~\cite{rau2016,winter2017}}. One prominent example is the Kitaev model that was initially studied in the context of quantum-spin-liquid states with anyonic excitations~\cite{hermanns2018} and can be realized in Ir$^{4+}$ or Ru$^{3+}$ compounds~\cite{jackeli2009}. Later work showed that {\cred additional terms in the spin Hamiltonian are often detrimental to these spin-liquid states, but give rise to a plethora of magnetically ordered states that are also highly unusual~\cite{winter2017,zhao2016,jain2017}.}

Here, we focus on $\beta$-Li$_2$IrO$_3$ that entails a hyperhoneycomb lattice of the Ir$^{4+}$ ions~\cite{biffin2014,takayama2015} {\cred and can be described by the $J-K-\Gamma$ model,
\begin{align}
 \hat H =\!\!\sum_{\langle ij\rangle; \alpha,\beta\neq \gamma} [J_{ij}\mathbf S_i\mathbf S_j &+ K_{ij}S_i^\gamma S_j^\gamma\pm \notag\\
 &\pm\Gamma_{ij}(S_i^\alpha S_j^\beta + S_i^\beta S_j^\alpha)],
\label{eq:ham}
\end{align}
where $J_{ij}$ stands for the isotropic (Heisenberg) exchange, $K_{ij}$ is the Kitaev exchange, and $\Gamma_{ij}$ is the off-diagonal exchange anisotropy.} Kitaev interactions $K$ are believed to be strong in $\beta$-Li$_2$IrO$_3$~\cite{kim2015,katukuri2016}, although $\Gamma$ may be of similar strength~\cite{kim2016,lee2016}. 

Experimentally, $\beta$-Li$_2$IrO$_3$ shows an incommensurate non-coplanar magnetic order below $T_N\simeq 38$\,K~\cite{biffin2014}. {\cred The nature of this order reflects strong Kitaev interactions~\cite{kimchi2015} that compete with the $\Gamma$ term~\cite{ducatman2018,stavropoulos2018}. This microscopic scenario renders $\beta$-Li$_2$IrO$_3$ different from the planar honeycomb iridates Na$_2$IrO$_3$ and $\alpha$-Li$_2$IrO$_3$~\cite{winter2017}, where third-neighbor Heisenberg exchange acts to stabilize the magnetic order, while the $\Gamma$ term is of minor importance. On the other hand, similarities to $\alpha$-RuCl$_3$ with its sizable $\Gamma$ term~\cite{winter2016,winter2017} may be expected.
}

{\cred Magnetic order in $\alpha$-RuCl$_3$ can be suppressed in the applied field~\cite{winter2017,sears2017,wolter2017,baek2017}. $\beta$-Li$_2$IrO$_3$ shows a somewhat similar behavior, as the field applied along the $b$ direction (\Hb) reduces the N\'eel temperature and leads to an apparent suppression of magnetic order above $H_c\simeq 2.8$\,T~\cite{ruiz2017}.} However, resonant x-ray scattering (RXS) reveals a more complex scenario. Instead of abruptly disappearing at $H_c$, the incommensurate order dwindles away as it is gradually replaced by the commensurate zigzag-type spin-spin correlations that become predominant above $H_c$~\cite{ruiz2017}. 

This rather exotic behavior was rationalized in Refs.~\onlinecite{ducatman2018,rousochatzakis2018} that proposed the incommensurate ($Q\neq 0$) and zigzag-type commensurate ($Q=0$)~\footnote{The zigzag order has an antiferromagnetic component in the $ac$ plane but shows the $Q=0$ periodicity, because there are several Ir atoms in the primitive cell of $\beta$-Li$_2$IrO$_3$.} orders to be two facets of the same, so-called $K$-state stabilized by the competing $K$ and $\Gamma$ interactions on the hyperhoneycomb lattice. The evolution of the magnetization and spin-spin correlations for \Hb\ {\cred is then not a breakdown of magnetic order toward a spin liquid,} but a continuous transformation between the two components of the same ordered state, as confirmed by the nearly constant RXS intensity as a function of the field~\cite{ruiz2017}. 

The evolution of $\beta$-Li$_2$IrO$_3$ in fields applied perpendicular to the $b$ direction was not characterized in detail apart from an observation that the magnetization grows much slower than for \Hb, and no $H_c$ is observed in this case up to at least 7\,T~\cite{ruiz2017}. {\cred In the following, we show that the field \Hpb\ has minor influence on $\beta$-Li$_2$IrO$_3$ indeed and does not break the $Q\neq 0$ incommensurate order. Moreover, we probe the field-induced state above $H_c$ for \Hb\ and juxtapose it with} the pressure-induced state of $\beta$-Li$_2$IrO$_3$~\cite{veiga2017}, where thermodynamic measurements and local probes detect the breakdown of the incommensurate order above 1.4\,GPa and the formation of a partially frozen spin liquid~\cite{majumder2018}, although these effects may also result from a structural dimerization~\cite{takayama2019} that occurs in the same pressure range at low temperatures~\cite{veiga2019}. We also use nuclear magnetic resonance (NMR) as a local probe of the field-induced state above $H_c$. We confirm that the spin-spin correlations emerging below 40\,K are clearly visible on the NMR time scale, and static magnetic fields develop upon cooling. We thus find no similarity to the pressure-induced state, where no static fields were observed~\cite{majumder2018}.

\section{Results}
\subsection{Crystal growth and characterization}
Single crystals were grown from separated educts~\cite{freund2016}. Elemental Li and Ir were placed, respectively, in the lower and upper parts of an alumina crucible that was heated in air to $T=1020$\,$^\circ$C within 5~hours, held at this temperature for roughly one week, and furnace-cooled. Single crystals with the dimensions of about 0.5\,mm along each side were collected from the alumina spikes placed in the middle of the crucible between the educts in order to provide a well-defined condensation point~\cite{freund2016}. Crystals of $\alpha$-Li$_2$IrO$_3$ and $\beta$-Li$_2$IrO$_3$ may grow simultaneously at this temperature, but they are easily distinguishable using x-ray diffraction (XRD) and magnetization measurements. 

\begin{figure}
\includegraphics{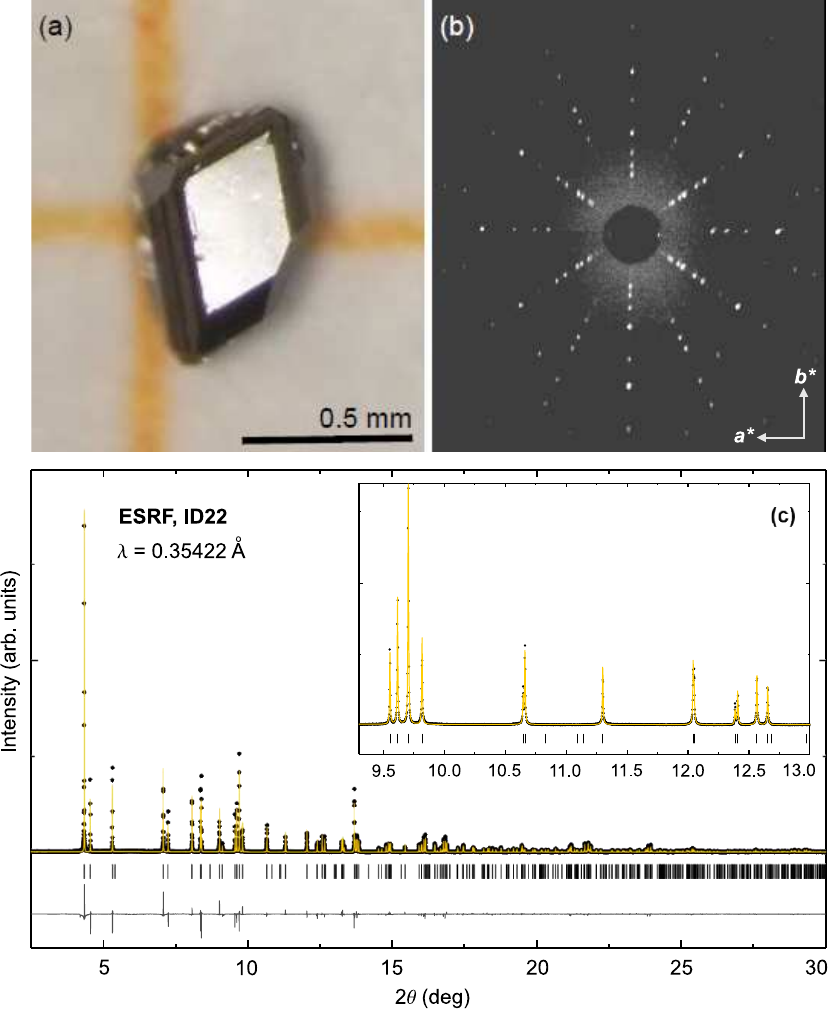}
\caption{\label{fig:growth}
(a) $\beta$-Li$_2$IrO$_3$ single crystal; (b) x-ray Laue-back-reflection-pattern for the beam parallel to [0\,0\,1];  (c) Rietveld refinement of the synchrotron XRD data: experimental (dots), calculated (yellow line), and difference patterns (gray line) are shown; the tick marks show the peak positions. The inset magnifies the pattern in the $2\theta=9.3-13.0^{\circ}$ range.
}
\end{figure}
A representative single crystal and its Laue-back-reflection pattern are shown in Fig.~\ref{fig:growth}. The Laue pattern was taken with a digital Dual FDI NTX camera manufactured by Photonic Science (tungsten anode, $U = 15$\,kV). The incident x-ray beam was oriented along [0\,0\,1], whereas the [1\,0\,0] and [0\,1\,0] directions in Fig.~\ref{fig:growth} are oriented roughly horizontally and vertically, respectively. 

The alignment was checked by measuring XRD from different crystal surfaces using the Rigaku Miniflex600 powder diffractometer (Cu-$K\alpha$ radiation). We also verified crystal quality and confirmed the absence of intergrowth phases, such as $\alpha$-Li$_2$IrO$_3$, by crushing several crystals from the same batch into powder and collecting high-resolution XRD data at the ID22 beamline of the European Synchrotron Radiation Facility (ESRF) at room temperature. The powder was placed into a thin-wall glass capillary and spun during the measurement. The diffracted signal was collected by nine scintillation detectors, each preceded by a Si (111) analyzer crystal yielding the instrumental peak broadening of about $0.004^{\circ}$ at $2\theta=10.5^{\circ}$. The reflections of \mbox{$\beta$-Li$_2$IrO$_3$} show a comparable full-width at half-maximum of about $0.006^{\circ}$ in this angular range (Fig.~\ref{fig:growth}c). No anisotropic peak broadening was observed, suggesting that the $\beta$-Li$_2$IrO$_3$ crystals are free from extended defects, such as staking faults that plagued the $\alpha$-Li$_2$IrO$_3$ samples~\cite{freund2016}.

\begin{table}
\caption{\label{tab:structure}
Fractional atomic coordinates ($x/a$, $y/b$, $z/c$) and atomic displacement parameters ($U_{\rm iso}$, in $10^{-2}$\,\r A$^2$) for $\beta$-Li$_2$IrO$_3$ obtained from the Rietveld refinement of the room-temperature powder XRD data collected at the ID22 beamline of the ESRF. The lattice parameters are $a=5.90648(2)$\,\r A, $b=8.45278(3)$\,\r A, and $c=17.8175(1)$\,\r A, and the space group is $Fddd$ (setting no. 2). The $U_{\rm iso}$ of oxygen were refined as a single parameter, the parameters for Li were fixed to those reported in Ref.~\onlinecite{takayama2015}. The error bars are from the Rietveld refinement. The refinement residuals are $R_I=0.053$ and $R_p=0.134$.
}
\begin{ruledtabular}
\begin{tabular}{cccccc}
 Atom & Site & $x/a$ & $y/b$ & $z/c$ & $U_{\rm iso}$ \\
 Ir & $16g$ & $\frac18$ & $\frac18$ & 0.70864(4) & 0.34(1)  \\
 O1 & $16e$ & 0.855(2)  & $\frac18$ & $\frac18$  & 0.42(7)  \\
 O2 & $32h$ & 0.636(2)  & 0.3631(6) & 0.0385(4)  & 0.42(7)  \\
 Li1 & $16g$ & $\frac18$ & $\frac18$ & 0.0498    & 0.5      \\
 Li2 & $16g$ & $\frac18$ & $\frac18$ & 0.8695    & 0.5      \\ 
\end{tabular}
\end{ruledtabular}
\end{table}

Structure refinement of the synchrotron data using the \texttt{Jana2006} software~\cite{jana2006} leads to the lattice parameters and atomic positions for Ir and O (Table~\ref{tab:structure}) in good agreement with the previous publications~\cite{biffin2014,takayama2015}. The parameters for Li showed large fluctuations and had to be fixed, given the large difference in the scattering powers of Li and Ir.

\subsection{Magnetization: temperature dependence}
Magnetization was measured {\cred on an individual 0.3\,mg single crystal} using the MPMS\,3 from Quantum Design in the temperature range of $1.8-400$\,K and in magnetic fields up to 7\,T. In higher fields up to 14\,T, the data were collected using the vibrating sample magnetometer option of Quantum Design PPMS. The crystal was weighed with a microgram balance and glued onto a quartz sample holder with a small amount of GE varnish that gives a negligible contribution to the signal even for the small crystals investigated in this work. 

\begin{figure}
{\centering {\includegraphics[width=10cm]{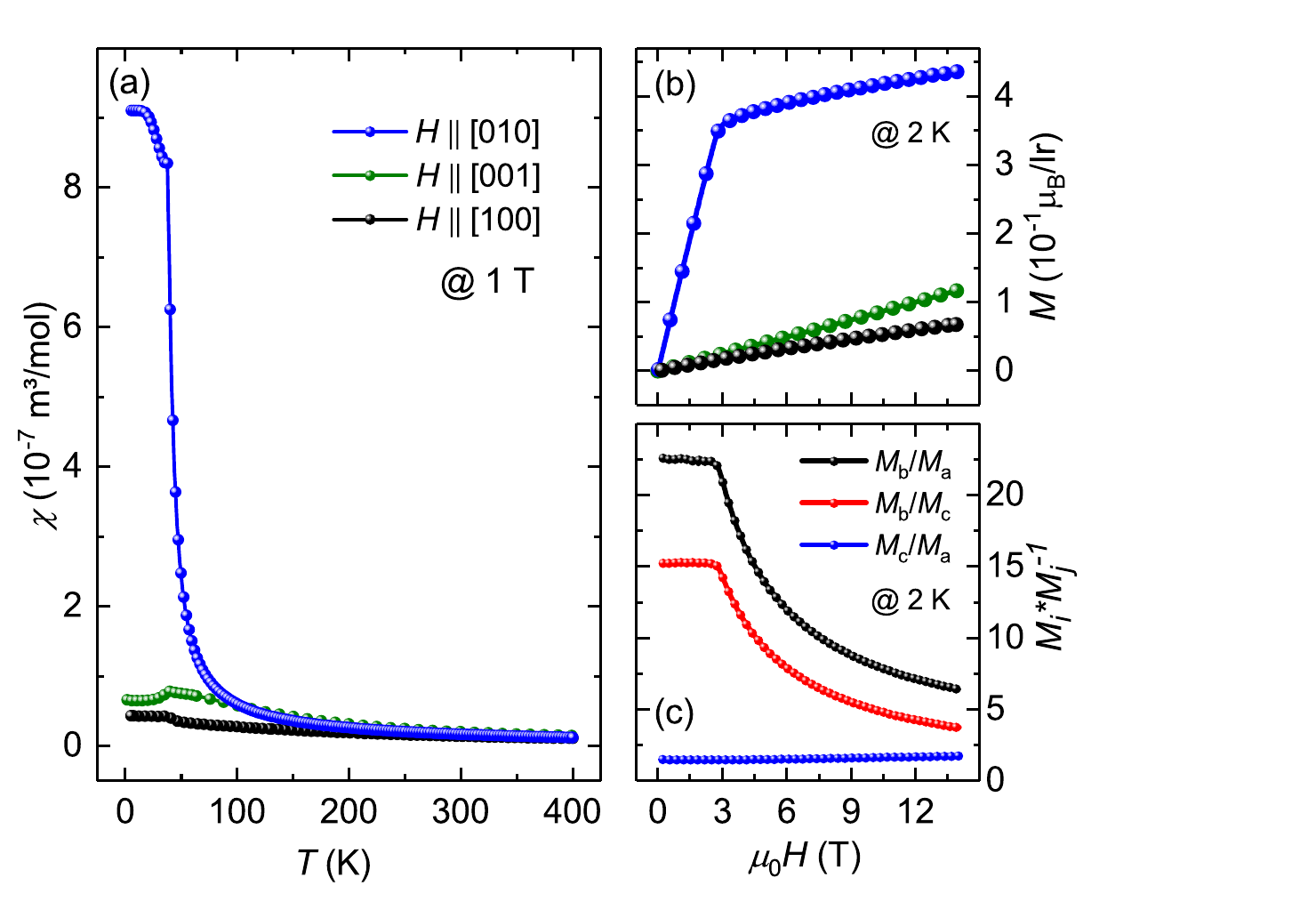}}\par} 
\caption{\label{fig:chivsT}
(a) Temperature dependence of the magnetic susceptibility $\chi$ measured on an individual single crystal at $H=1$\,T for three field directions. (b) Field dependence of the magnetization measured along the same directions up to 14\,T at 2\,K. (c) Anisotropy of the magnetization as a function of field.
} 
\end{figure}

\begin{figure}
{\centering {\includegraphics[width=10cm]{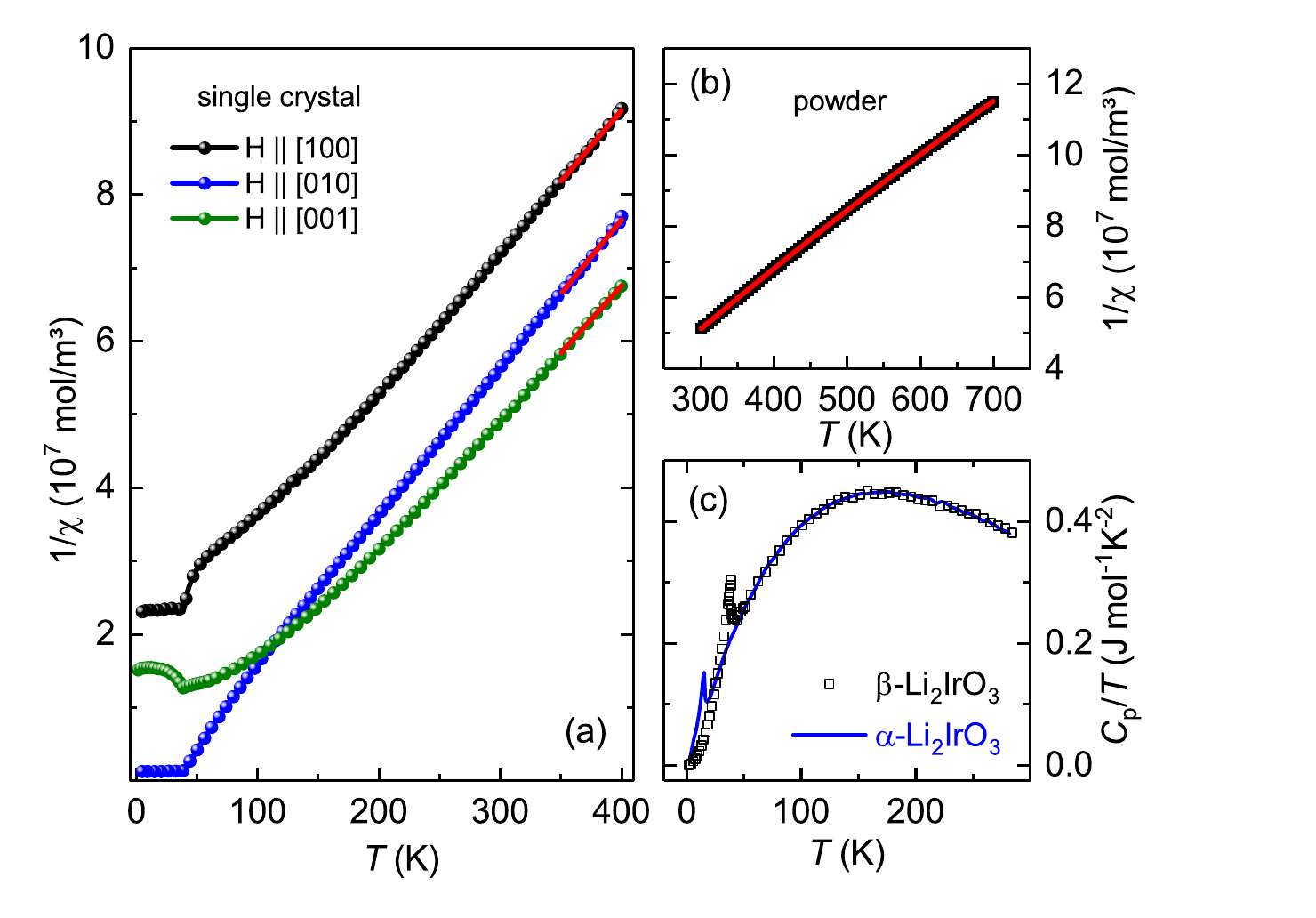}}\par} 
\caption{\label{fig:CWfit}
(a) Inverse susceptibility measured on an individual crystal in the field $H=1$\,T applied along three different directions; the solid lines are examples of the Curie-Weiss fits in the $350-400$\,K range. (b) Inverse susceptibility measured above 300\,K on a powder sample ($H=1$\,T); the solid line is the Curie-Weiss fit. (c) Zero-field specific heat of $\alpha$-Li$_2$IrO$_3$~\cite{freund2016} and $\beta$-Li$_2$IrO$_3$ (this work) measured on stacks of single crystals.
} 
\end{figure}

\begin{figure*}
{\centering {\includegraphics[width=17cm]{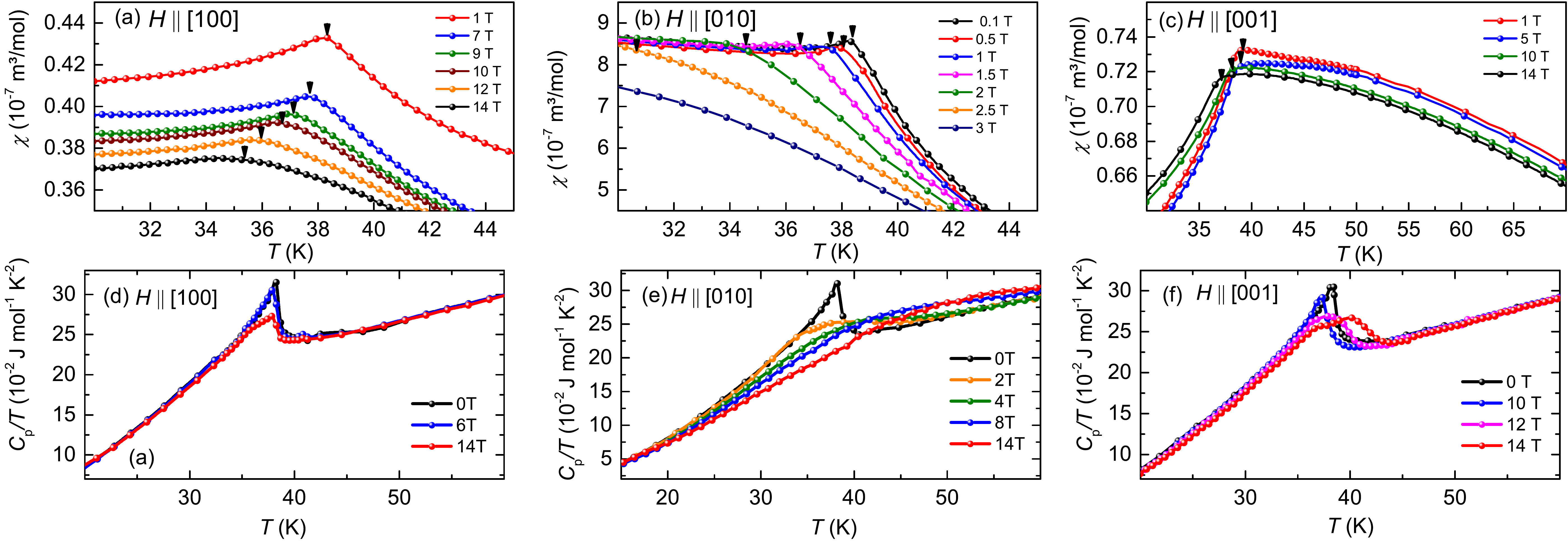}}\par} 
\caption{\label{fig:chiCp}
(a), (b) and (c) Temperature dependence of the magnetic susceptibility at different magnetic fields applied along the $a$, $b$, and $c$ directions, respectively. (d), (e) and (f) Temperature dependence of $C_p/T$ at different magnetic fields applied along the $a$, $b$ and $c$ directions, respectively.
} 
\end{figure*}

{\cred The field of 1\,T was chosen for temperature-dependent susceptibility measurements to ensure a large enough signal, especially at high temperatures. In this magnetic field, no difference between field-cooled and zero-field-cooled regimes was observed.} 
The susceptibility ($\chi$) shows a clear transition anomaly at $T_N\simeq 38$\,K for all field directions (Figs.~\ref{fig:chivsT} and~\ref{fig:chiCp}). At higher temperatures, linear behavior of the inverse susceptibility (Fig.~\ref{fig:CWfit}) signals the Curie-Weiss regime $\chi=\chi_0+C/(T-\Theta)$. {\cred However, the ensuing parameters strongly depend on the temperature range of the fit. At first glance, the data above 100\,K could be used, because above this temperature specific heat of $\beta$-Li$_2$IrO$_3$ becomes nearly indistinguishable from that of $\alpha$-Li$_2$IrO$_3$ (Fig.~\ref{fig:CWfit}c), suggesting that most of the magnetic entropy is released below 100\,K. On the other hand, inverse susceptibility remains non-linear up to $200-250$\,K (Fig.~\ref{fig:CWfit}a).}

{\cred To check whether this non-linearity arises from the temperature-independent $\chi_0$ term, we first performed susceptibility measurements above 400\,K using the oven option of the MPMS. Individual crystals proved too small for such a measurement, so we used a powder sample that was sealed into a quartz ampoule. Background from the ampoule and oven was subtracted. The fit to the resulting high-$T$ data in the $300-700$\,K range yields the temperature-independent contribution $\chi_0=1.1\times 10^{-9}$\,m$^3$/mol (Fig.~\ref{fig:CWfit}b). The positive $\chi_0$ contribution leads to a weak curvature of $1/\chi$ above 600\,K and can not account for the more pronounced downward curvature below 250\,K.}

In our case, $\chi_0$ stands for a combination of two temperature-independent contributions, the negative one from the core diamagnetism estimated as $\chi_{\rm core}=-8.41\times 10^{-10}$\,m$^3$/mol~\cite{bain2008}, and the positive one from the van Vleck paramagnetism, $\chi_{\rm VV}$. {\cred Using $\chi_0=\chi_{\rm core}+\chi_{\rm VV}$, we estimate $\chi_{\rm VV}=1.94\times 10^{-9}$\,m$^3$/mol that is comparable to $1.3\times 10^{-9}$\,m$^3$/mol (Na$_2$IrO$_3$~\cite{mehlawat2017}), $2.7\times 10^{-9}$\,m$^3$/mol ($\alpha$-Li$_2$IrO$_3$~\cite{mehlawat2017}), and $1.4\times 10^{-9}$\,m$^3$/mol (K$_2$IrCl$_6$~\cite{khan2019}) reported in the previous literature for Ir$^{4+}$ in the $j_{\rm eff}=\frac12$ state.}

{\cred We now fix $\chi_0$ to the value obtained above, and keep only $C$ and $\Theta$ as the fitting parameters for the single-crystal data. The fits are performed between $T_{\min}$ and 400\,K, where $T_{\min}=250$, 300, and 350\,K. The average of the three values and their spread were taken as the best estimate and the error bar for the fitting parameter, respectively. The resulting Curie-Weiss temperatures $\Theta$ and paramagnetic effective moments $\mu_{\rm eff}$ extracted from the Curie constants $C$ for different field directions are summarized in Table~\ref{tab:CW}.}

\begin{table}
\caption{\label{tab:CW}
Results of the Curie-Weiss fitting for different field directions. The error bars are obtained from fitting the data in different temperature ranges as explained in the text. The temperature-independent contribution $\chi_0=1.1\times 10^{-9}$\,m$^3$/mol was kept fixed in the fit.
}
\begin{ruledtabular}
\begin{tabular}{c@{\hspace{1.5cm}}rc}
Direction & $\Theta$\,(K) &  $\mu_{\rm eff}$\,($\mu_B$) \\
\Ha & $-33(3)$ & 1.64(1) \\
\Hb &   44(4)  & 1.65(2) \\
\Hc &   53(4)  & 1.74(2) \\ 
\end{tabular}
\end{ruledtabular}
\end{table}

{\cred The effective moments are rather isotropic and only slightly deviate from 1.73\,$\mu_B$ expected for Ir$^{4+}$ in the $j_{\rm eff}=\frac12$ state. This is well in line with the earlier \textit{ab initio} results~\cite{kim2015,katukuri2016} that suggested the applicability of the \mbox{$j_{\rm eff}=\frac12$} scenario to $\beta$-Li$_2$IrO$_3$. The Curie-Weiss temperatures demonstrate a sizable anisotropy with $\Theta_a<\Theta_b<\Theta_c$. We also note that our Curie-Weiss parameters are somewhat different from those reported in the Supplemental Material of Ref.~\onlinecite{ruiz2017}, where the effective moments exceed 1.73\,$\mu_B$ for all field directions, and $\Theta_a<\Theta_c<\Theta_b$. This discrepancy is likely related to the fact that in Ref.~\onlinecite{ruiz2017} the data above 100\,K were used for the Curie-Weiss fit, and no $\chi_0$ term was included. 
}

\begin{figure}
\includegraphics[width=8cm]{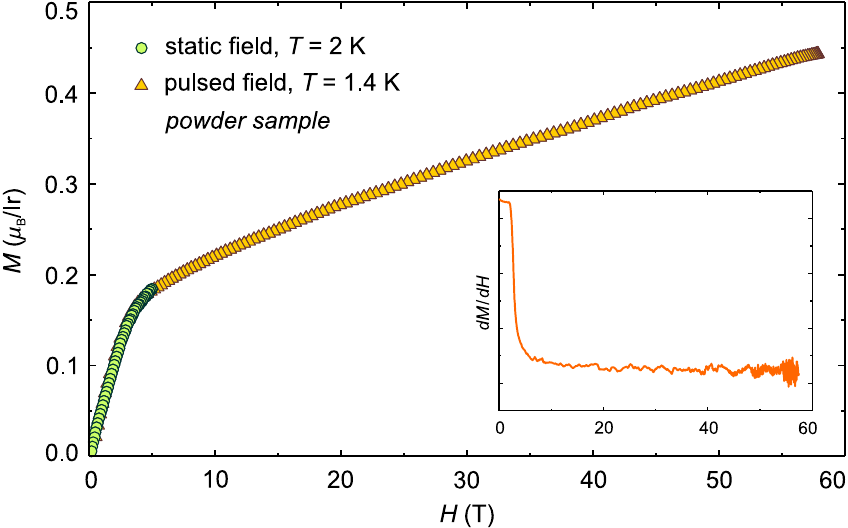}
\caption{\label{fig:hld}
Field-dependent magnetization measured on a powder sample in static and pulsed fields at $T=2$\,K and 1.4\,K, respectively. The inset shows the derivative of the pulsed-field data.
} 
\end{figure}

\begin{figure}
{\centering {\includegraphics[width=8.5cm]{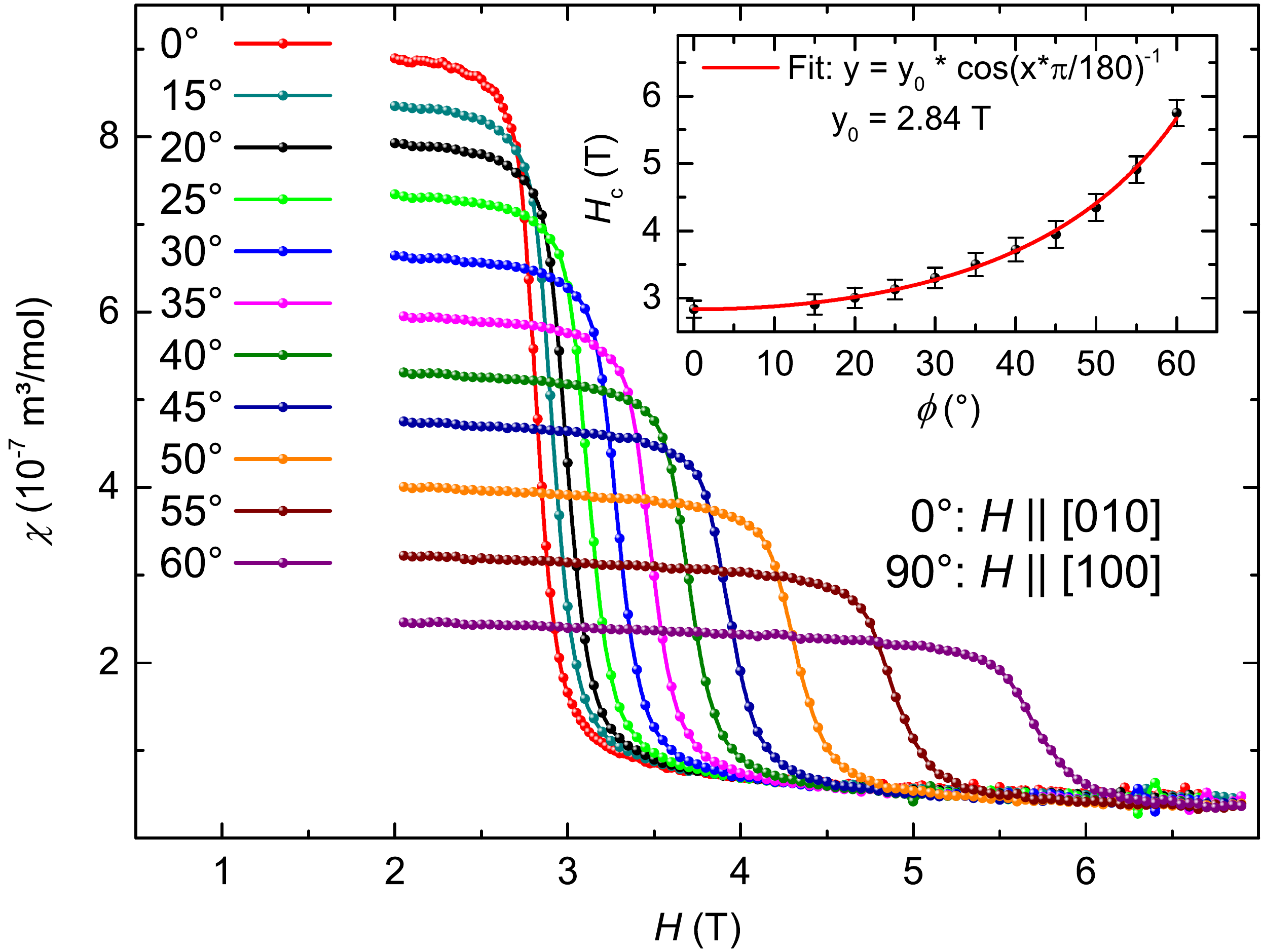}}\par} 
\caption{\label{fig:chivsH}
Field-dependent magnetic susceptibility measured in the fields applied in between the $a$ and $b$ directions. The transition field $H_c^{\phi}$ is defined as the midpoint of the step-like decrease in $\chi$. The inset shows the angular dependence of $H_c^{\phi}$ and its cosine fit as explained in the text.
} 
\end{figure}

\subsection{Magnetization: field dependence}
Magnetization of $\beta$-Li$_2$IrO$_3$ is strongly anisotropic not only as a function of temperature but also as a function of field (Fig.~\ref{fig:chivsT}a). We observe a fast increase in the magnetization for \Hb\ and a much lower slope of $M(H)$ for the two other directions. The kink is detected at $H_c\simeq 2.8$\,T for \Hb, whereas no kink is seen for \Ha\ and \Hc\ up to 14\,T. The magnetization at $H_c$ is very close to $\frac13$ of the saturation value (1\,$\mu_B$/f.u.) expected for Ir$^{4+}$ with $j_{\rm eff}=\frac12$. It is in good agreement with the previous reports~\cite{takayama2015,ruiz2017}, although we note that the data of Ref.~\onlinecite{takayama2015} were apparently taken on a single crystal or at least on a well-aligned powder sample, whereas powder samples with random crystallite orientations show a smeared kink at $H_c$ with the much lower $M(H_c)\simeq 0.15$\,$\mu_B$/f.u.~\cite{majumder2018} (see also Fig.~\ref{fig:hld}).

To probe the magnetization in higher fields, we measured the powder sample of $\beta$-Li$_2$IrO$_3$ (single crystals were too small for this measurement) using the pulsed-field setup at the High Magnetic Field Laboratory in Dresden. The sample was loaded into a teflon tube and placed into the magnet that yields fields up to 58\,T with a rise time of 7\,ms and the total pulse duration of about 20\,ms. Details of the measurement procedure have been described elsewhere~\cite{tsirlin2009}. The data in Fig.~\ref{fig:hld} demonstrate the linear increase in $M(H)$ above $H_c$, as confirmed by the flat $dM/dH$ curve. This suggests the absence of any further field-induced transformations above $H_c$ within the resolution of our measurement.

The kink at $H_c$ is solely caused by \Hb. We demonstrate this by field-dependent measurements for different directions of the applied field that vary between the $a$ and $b$ axes. An abrupt step-like change in the susceptibility typical of a second-order phase transition was observed (Fig.~\ref{fig:chivsH}). The mid-points define the transition field $H_c^{\phi}$ that follows a simple cosine function $H_c^{\phi}=H_c\cos\phi$ with $H_c=2.84(1)$\,T. This observation implies that the field-induced state is triggered by the projection of the field on the $b$-axis, whereas the $a$-component of the magnetic field remains inactive. 

\subsection{Specific heat and phase diagram}
\begin{figure}
{\centering {\includegraphics[width=8.5cm]{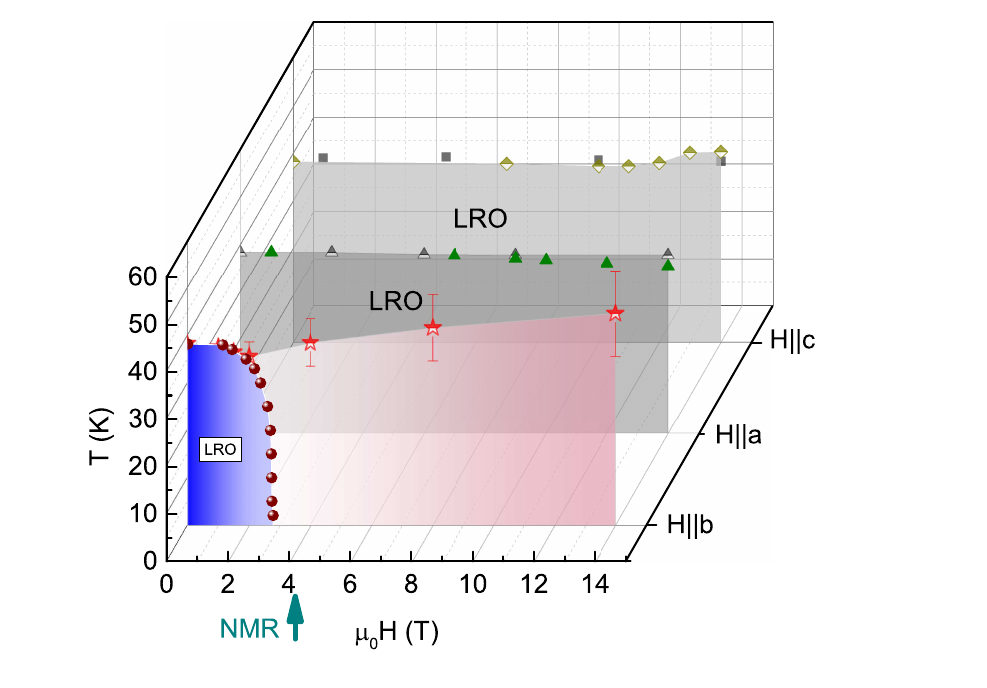}}\par} 
\caption{\label{fig:phasediagram}
Temperature-field phase diagram obtained from the magnetization (filled symbols) and specific heat (half-filled symbols) data collected for three directions of the applied field. The LRO (long-range order) stands for the region where a phase transition is observed as a function of temperature. According to Ref.~\onlinecite{ruiz2017}, this region is characterized by the presence of $Q\neq 0$ spin-spin correlations that break the symmetry and produce a distinct ordered phase separated from the paramagnetic state by a phase transition. For \Hb\ above $H_c$, only non-symmetry-breaking $Q=0$ correlations are present, leading to a crossover denoted by stars. The arrow indicates the field around which the $^7$Li NMR measurements were performed.
} 
\end{figure}

To determine specific heat, we assembled mosaics of several co-aligned single crystals and performed the measurement in the temperature range of $1.8-100$\,K and field range up to 14\,T in Quantum Design PPMS using the relaxation method. A sharp $\lambda$-type anomaly is observed in zero field. For \Ha\ and \Hc\ the anomaly retains its shape and shifts with the field only marginally (Fig.~\ref{fig:chiCp}). Above 10\,T, the anomaly broadens {\cred and may even split into two, but we attribute this effect to a slight misalignment of the crystals in the mosaic, because magnetization measured on an individual single crystal (Fig.~\ref{fig:phasediagram}c) still shows one transition only.} In contrast, the field applied along $b$ blurs the anomaly already at 2\,T. Above $H_c$, the anomaly disappears, giving way to a broad hump that shifts toward higher temperatures upon increasing the field. 

{\cred At higher temperatures, specific heat of $\beta$-Li$_2$IrO$_3$ is dominated by the phonon contribution. Fig.~\ref{fig:CWfit}c shows that above 70\,K specific heats of $\alpha$- and $\beta$-Li$_2$IrO$_3$ are nearly indistinguishable. Attempts to separate the magnetic and phonon contributions were so far unsuccessful given the absence of a suitable phonon reference for either of the Li$_2$IrO$_3$ polymorphs.
}

By combining the specific-heat and magnetization data, we construct a $T-H$ phase diagram for three directions of the applied field (Fig.~\ref{fig:phasediagram}). {\cred Transition temperatures are determined from the peak positions in $C_p(T)$ and $d\chi/dT$, respectively.} For \Ha\ and \Hc\ the transition temperature decreases and reaches about 35.5\,K at 14\,T. For \Hb\ the transition anomaly in the specific heat becomes too broad already in low field, so it is more convenient to track the phase boundary using field-dependent magnetization. Above $H_c$, the field-induced phase does not show any transition as a function of temperature, suggesting that the formation of the $Q=0$ zigzag-type correlations is only a crossover, similar to the onset of magnetization in ferromagnets~\cite{ruiz2017}. The crossover temperature can be tracked by the position of the hump in the specific heat, which we also show on the phase diagram.

\subsection{$^7$Li NMR spectra and line shift} 
NMR experiments require larger samples, so we assembled a mosaic of about 20 single crystals that were co-aligned along the $b$ direction facilitating the measurements for either \Hb\ or \Hpb. The measurements were performed at the fixed frequency of 70\,MHz and field-sweep spectra have been taken using a conventional pulsed NMR technique. The field strength of 4.23\,T places the system into the field-induced state for \Hb, but leaves it in the incommensurately ordered state for \Hpb\ (Fig.~\ref{fig:phasediagram}). 

\begin{figure}
{\centering {\includegraphics[width=8.5cm]{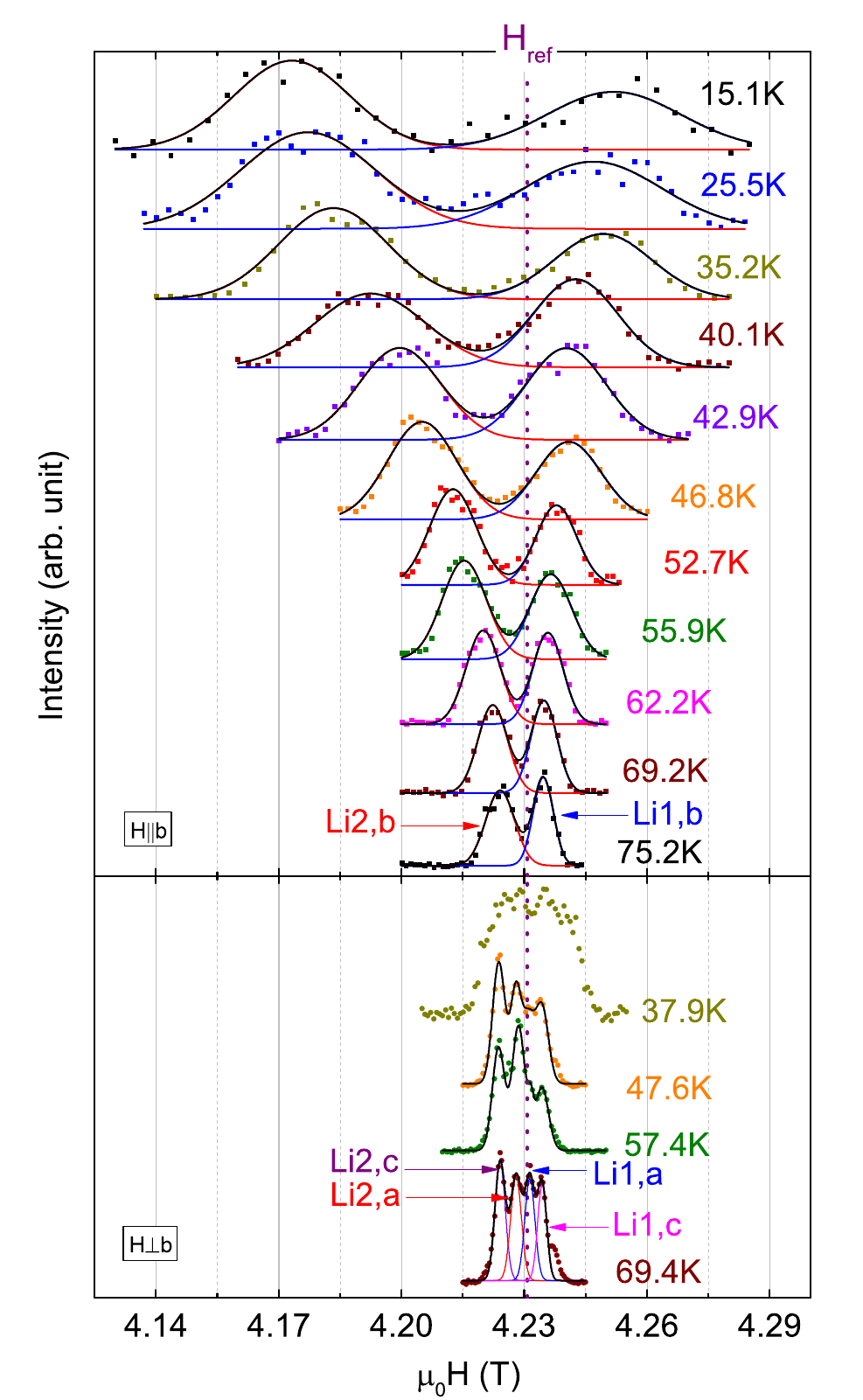}}\par} 
\caption{\label{fig:spectra}
Field-sweep $^7$Li NMR spectra measured at a fixed frequency of 70\,MHz for \Hb\ (upper panel) and \Hpb\ (lower panel). The dotted vertical line indicates the reference field for 70\,MHz.
} 
\end{figure}

\begin{figure*}
{\centering {\includegraphics[width=17.5cm]{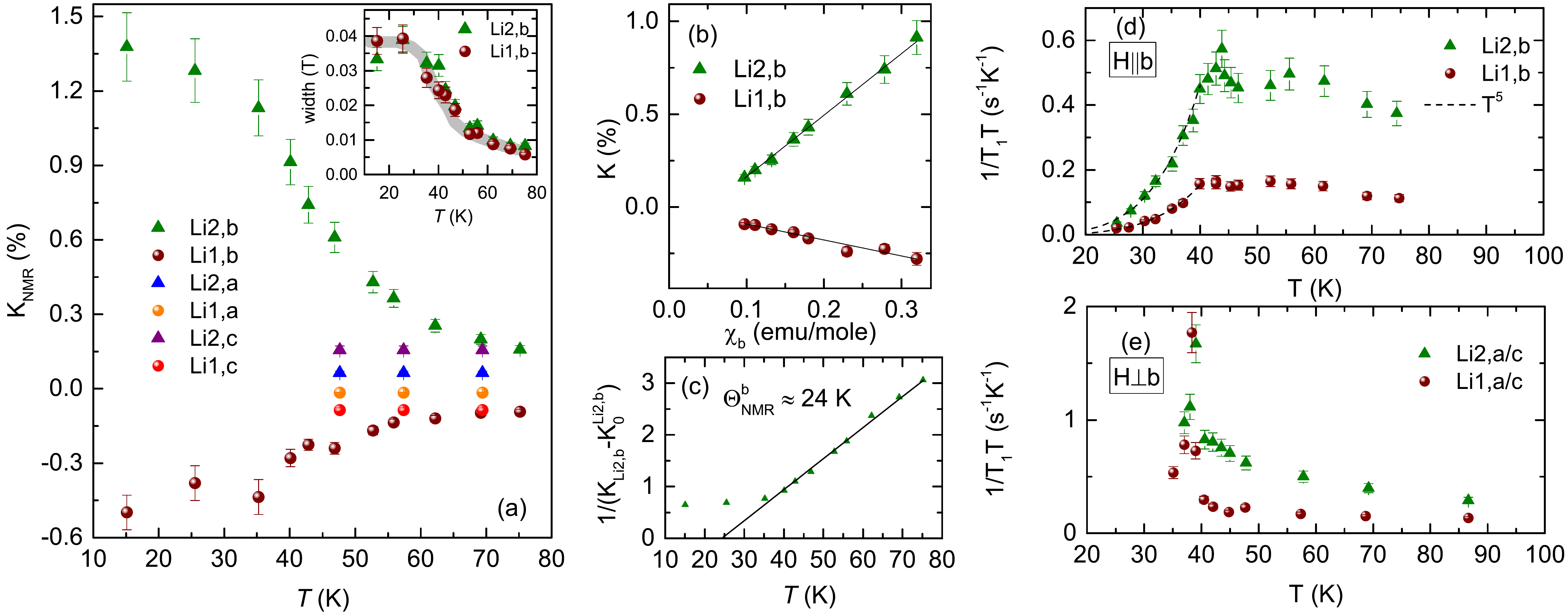}}\par} \caption{\label{fig:NMRpara}(a): Temperature dependence of the line shift ($K_{\rm NMR}$) for different Li sites and field directions. The inset shows temperature evolution of the line width for \Hb. (b) $K_{\rm NMR}$ versus $\chi$ for the Li1 and Li2 sites and \Hb. (c) Inverse of $(K^{{\rm Li2},b}-K_0^{{\rm Li2},b})$ as a function of temperature, with the solid line showing the linear Curie-Weiss fit above $T_N$. (d)-(e) Temperature dependence of $1/T_1T$ for \Hb\ (d) and \Hpb\ (e).} 
\end{figure*}

Figure~\ref{fig:spectra} shows temperature dependence of the field-sweep $^7$Li ($I=\frac32$) NMR spectra. Two different crystallographic sites of Li (Table~\ref{tab:structure}) are expected to probe different transferred hyperfine fields from the surrounding magnetic Ir$^{4+}$ ions. The Li1 atoms have four Li--Ir contacts of about 3.0\,\r A, all mediated by oxygen, whereas the Li2 atoms reveal five such contacts and may experience stronger hyperfine fields, resulting in a stronger temperature dependence of the NMR line shift $K$ and in larger values of the spin-lattice relaxation rates $1/T_1$. 

The assignment of two distinct spectral lines to the Li1 and Li2 sites is shown in Fig.~\ref{fig:spectra} (top frame) for \Hb. For the perpendicular orientation \Hpb, we expect two pairs of spectral lines corresponding to \Ha\ and \Hc, respectively. Four spectral lines are observed indeed (Fig.~\ref{fig:spectra}, bottom frame). At high temperatures, the area under each of these lines for a given field direction yields the intensity ratios around 1:1 in agreement with the equal abundance of the Li1 and Li2 sites in the crystal structure (Table~\ref{tab:structure}). It is also worth mentioning that none of the observed $^7$Li NMR spectral lines exhibit the quadrupolar splitting. The sharp lines, especially for \Hpb, confirm the crystal quality and the low defect concentration in agreement with our XRD results (Fig.~\ref{fig:growth}), whereas the absence of the quadrupolar splitting simplifies determination of the NMR parameters, unlike in the case of $^{35}$Cl NMR in $\alpha$-RuCl$_3$, where strong quadrupolar splitting had to be overcome by applying the field along special directions that did not match the crystallographic ones~\cite{baek2017}.

Temperature dependence of the line shift for different field directions and for both Li sites is plotted in Fig.~\ref{fig:NMRpara}(a). In the \Hpb\ case, no appreciable temperature dependence was observed suggesting that either the hyperfine coupling $A_{\rm hf,\perp}$ is small, or weak changes in the bulk susceptibility (Fig.~\ref{fig:chivsT}) are not sufficient to cause a significant change in $K$. On the other hand, a strong temperature dependence can be seen for \Hb. The $A_{\rm hf,\|}$ in this case was determined from the relation $K_{\rm NMR} = K_0+ (A_{\rm hf}/N\mu_B)\chi$, where $K_0$ is the temperature-independent contribution, and $\chi$ is the bulk magnetic susceptibility. The slope of the linear $K_{\rm NMR}-\chi$ relation (Fig.~\ref{fig:NMRpara}b) yields $A_{\rm hf,\|}=-0.047$\,kOe/$\mu_B$, $K_0=0$\% for Li1 and $A_{\rm hf,\|}=0.18$\,kOe/$\mu_B$, $K_0=-0.168$\% for Li2. By subtracting $K_0$, we obtain a local measure of $\chi$ that follows the Curie-Weiss behavior above $T_N$ (Fig.~\ref{fig:NMRpara}c). The extracted Curie-Weiss temperature of 24\,K is in reasonable agreement with the bulk value for \Hb\ (Table~\ref{tab:CW}).

Temperature evolution of the NMR linewidth shows a strongly anisotropic behavior too. In the \Hpb\ case, the linewidth is nearly temperature-independent down to $T_N$, where the spectrum broadens abruptly due to the development of inhomogeneous local fields in the magnetically ordered state (Fig.~\ref{fig:spectra}). The \Hb\ spectra show instead a more gradual increase in the line broadening (see the inset to Fig.~\ref{fig:NMRpara}a). The faster increase in the linewidth in the $30-50$\,K temperature range indicates the onset of spin-spin correlations, but the overall behavior is reminiscent of a gradual crossover suggested also by the specific heat data above $H_c$ (Fig.~\ref{fig:chiCp}e).

\subsection{$^7$Li NMR spin-lattice relaxation rate}
To obtain the spin-lattice relaxation rate $1/T_1$, we measured the magnetization recovery and fitted it by a single exponential function. Such fits were possible within the entire temperature range of our measurement. The absence of stretched-exponential behavior indicates the homogeneity of the magnetic state. For \Hpb\ we were unable to separately measure $1/T_1$ for \Ha\ and \Hc, so these data show the cumulative response from both field directions.

In general~\cite{moriya1962},
\begin{equation}
(1/T_1T)_{H\|\,\alpha} \propto \sum_{q,\omega_n\rightarrow 0} A_{{\rm hf},\perp \alpha}^2(q)\times \frac{\chi_{\perp \alpha}'' (q,\omega_n)}{\omega_n},
\end{equation}
where $\chi''$ is the imaginary part of the dynamic spin susceptibility, $\omega_n$ is the nuclear Larmor frequency, and we choose an arbitrary field direction $\alpha$. Assuming the similar magnetic response for \Ha\ and \Hc, we can restrict the problem to the parallel and perpendicular components of $A_{\rm hf}$ and, likewise, of $\chi''$. Then for \Hb,
\begin{equation}
(1/T_1T)_{H\|b} \propto \sum_{q,\omega_n\rightarrow 0} \left[ 2A_{\rm hf,\perp}^2(q)\times\frac{\chi_\perp''(q,\omega_n)}{\omega_n} \right],
\end{equation}
whereas for the perpendicular field direction
\begin{align}
& (1/T_1T)_{H\perp b} \propto \notag\\[5pt] 
 & \sum_{q,\omega_n\rightarrow 0} \left[A_{\rm hf,\|}^2(q) \frac{\chi_{\|}''(q,\omega_n)}{\omega_n} + A_{\rm hf,\perp}^2(q) \frac{\chi_\perp''(q,\omega_n)}{\omega_n} \right].
\end{align}
Experimentally, in the paramagnetic state above 50\,K similar values of $(1/T_1T)$ are observed in both cases (Fig.~\ref{fig:NMRpara}de). Given that $A_{\rm hf,\|}>A_{\rm hf,\perp}$, this implies $\chi_\perp''(q,\omega_n)>\chi_\parallel''(q,\omega_n)$.

Below 40\,K, $1/T_1T$ follows a power-law behavior $T^\nu$ with $\nu\approx 5$. The exponents $\nu=4$ and 2 are expected if temperature exceeds the magnon gap $\Delta$, and the nuclear spin-lattice relaxation is governed by a three-magnon Raman process or by a two-magnon Raman process, respectively. The faster decrease in $1/T_1T$ upon cooling may indicate that the excitation gap is comparable in size to the measurement temperature. 

\section{Discussion and Summary}
The magnetic response of $\beta$-Li$_2$IrO$_3$ is strongly anisotropic. At high temperatures, the anisotropy of magnetic susceptibility (Fig.~\ref{fig:chivsT}) manifests itself by the different Curie-Weiss temperatures, as in the planar honeycomb iridates where paramagnetic effective moments~\cite{winter2017} are close to 1.73\,$\mu_B$ expected for the $j_{\rm eff}=\frac12$ state of Ir$^{4+}$, while the Curie-Weiss temperatures vary by more than 100\,K depending on the field direction~\cite{winter2017}. In $\beta$-Li$_2$IrO$_3$, the $\Theta$ values are clearly shifted to the ferromagnetic side, as noticed from a comparison between the powder-averaged $\Theta_{\rm av}$ of $-127$\,K for Na$_2$IrO$_3$~\cite{mehlawat2017}, $-105$\,K for $\alpha$-Li$_2$IrO$_3$~\cite{mehlawat2017}, and $+40$\,K for $\beta$-Li$_2$IrO$_3$~\cite{takayama2015}. {\cred Our present estimate of $\Theta_{\rm av}=21$\,K confirms this trend.}

{\cred The Curie-Weiss temperatures for different directions of the applied field are calculated as
\begin{equation}
 \Theta_{\alpha}=-\frac14\sum_{\langle ij\rangle}\,\mathbf h_{\alpha}^{\dagger}\,\mathbb J_{ij}\,\mathbf h_{\alpha},
\end{equation}
where $\mathbb J_{ij}$ are the exchange tensors, and $\mathbf h_{\alpha}$ is a unitary vector in the direction of the field. Each Ir$^{4+}$ ion forms three exchange bonds with its nearest neighbors. These bonds are designated by $X$, $Y$, and $Z$ depending on the direction of the Kitaev term. Similar to Refs.~\onlinecite{ducatman2018} and~\onlinecite{majumder2018}, we use $\mathbf X=(\mathbf a+\mathbf c)/\sqrt2$, $\mathbf Y=(\mathbf c-\mathbf a)/\sqrt2$, and $\mathbf Z=-\mathbf b$, where $\mathbf a$, $\mathbf b$, and $\mathbf c$ stand for unit vectors along the respective crystallographic directions.}

{\cred In the $XYZ$ coordinate frame, exchange tensors take the form
\begin{gather*}
 \mathbb J_X=\left(
 \begin{array}{ccc}
  J+K & 0 & 0 \\
	0 & J & \Gamma \\
	0 & \Gamma & J
 \end{array}
 \right),\,\,\, 
 \mathbb J_Y=\left(
 \begin{array}{ccc}
  J & 0 & \Gamma \\
	0 & J+K & 0 \\
	\Gamma & 0 & J
 \end{array}
 \right), \\[5pt]
 \mathbb J_Z=\left(
 \begin{array}{ccc}
  J & \Gamma & 0 \\
	\Gamma & J & 0 \\
	0 & 0 & J+K
 \end{array}
 \right).
\end{gather*}
In the same coordinate frame, field directions are defined by
\begin{equation*}
 \mathbf h_a=\frac{1}{\sqrt 2}\left(
 \begin{array}{r}
  1 \\ -1 \\ 0
 \end{array}
 \right),\,\,
 \mathbf h_b=\frac{1}{\sqrt 2}\left(
 \begin{array}{c}
  0 \\ 0 \\ 1
 \end{array}
 \right),\,\,
 \mathbf h_c=\frac{1}{\sqrt 2}\left(
 \begin{array}{r}
  1 \\ 1 \\ 0
 \end{array}
 \right).
\end{equation*}
Then the Curie-Weiss temperatures are obtained as
\begin{align}
 \Theta_a & =-(3J+K-\Gamma)/4, \\
 \Theta_b & =-(3J+K)/4, \\
 \Theta_c & =-(3J+K+\Gamma)/4.
\label{eq:CW}
\end{align}
}

{\cred The Curie-Weiss temperatures listed in Table~\ref{tab:CW} reveal that $\Theta_a<\Theta_b<\Theta_c$ indeed. The combination of the Heisenberg and Kitaev terms, $3J+K$, can be estimated as $-4\Theta_b$ or $-2(\Theta_a+\Theta_c)$ resulting in $3J+K=-130\pm50$\,K. Likewise, we find $\Gamma=-170\pm 130$\,K. The large error bars reflect the fact that experimentally $\Theta_b\neq (\Theta_a+\Theta_c)/2$. This may be a drawback of the Curie-Weiss fitting performed in the limited temperature range, or an indication that the $J-K-\Gamma$ model does not fully capture the behavior of $\beta$-Li$_2$IrO$_3$. In Eq.~\eqref{eq:ham}, we assumed same values of $J$, $K$, and $\Gamma$ on all bonds, but the $X$- and $Y$- bonds are not related to the $Z$-bonds by symmetry and may feature different exchange parameters. At this point, we can only conclude that our Curie-Weiss parameters are consistent with the general microscopic scenario of $K<0$ and $\Gamma<0$ implied by the recent theory studies~\cite{ducatman2018,rousochatzakis2018,stavropoulos2018}. Further refinement of the interaction parameters would require additional experimental input and goes beyond the scope of our present manuscript.
}

Turning now to the low-temperature anisotropy, we recognize that it is quite different from the high-temperature one. The $b$-direction is singled out, whereas similar magnetic response is observed for \Ha\ and \Hc. The \Hb\ regime leads to a kink in the magnetization accompanied by the suppression of $T_N$. The two other field directions cause only a marginal reduction in the $T_N$ (Fig.~\ref{fig:phasediagram}), with no field-induced transitions observed up to at least 58\,T. The origin of this anisotropy lies not in the model itself, but in the symmetry of the magnetically ordered state ($K$-state) that combines the $Q\neq 0$ and $Q=0$ components~\cite{rousochatzakis2018}. The latter component couples to the field applied along $b$ and, most importantly, to the longitudinal magnetization caused by this field~\cite{ducatman2018}. This unusual mechanism leads to a very fast suppression of the $Q\neq 0$ order and, consequently, of the $T_N$. On the other hand, \Ha\ and \Hc\ lack the benefit of such a coupling and will polarize the system only after they overcome $\Gamma$, which is the leading term of the order of 100\,K~\footnote{The estimates of this term vary between about 40\,K~\cite{katukuri2016} and 150\,K~\cite{majumder2018}, but even the lowest value is likely large enough to prevent any field-induced transitions for fields up to 58\,T.}.

From the purely thermodynamic perspective, the zero-field transition at $T_N$ resembles the second-order transition, as a sharp $\lambda$-type anomaly is observed in the specific heat (Fig.~\ref{fig:chiCp}) and thermal expansion~\cite{majumder2018}. This transition remains second-order also in the applied field, in agreement with the symmetry analysis of Ref.~\onlinecite{ruiz2017}. However, above $H_c$ the transition disappears, because the $Q\neq 0$ mode is fully suppressed, whereas the remaining $Q=0$ mode does not lift any symmetry and appears as a crossover between the paramagnetic and partially polarized (quantum paramagnetic) states~\cite{ruiz2017}.

The evolution of $\beta$-Li$_2$IrO$_3$ for \Hb\ bears certain similarities to the behavior of $\alpha$-RuCl$_3$ under in-plane magnetic fields~\cite{winter2017}. In both cases, thermodynamic phase transition is suppressed as $Q\neq 0$ spin-spin correlations give way to the $Q=0$ correlations~\cite{wolter2017,sears2017,banerjee2018,winter2018}. Moreover, the NMR response of $\beta$-Li$_2$IrO$_3$ in the field-induced state at 4.3\,T looks similar to the response of $\alpha$-RuCl$_3$ above 9\,T~\cite{baek2017} with the gradual development of local fields and a maximum in $1/T_1T$. In $\alpha$-RuCl$_3$, the presence of an intermediate spin-liquid phase around $H_c=7$\,T is presently debated~\cite{kelley2019,balz2019}, but such a phase is clearly absent in $\beta$-Li$_2$IrO$_3$, where we observe a single field-induced transition (Fig.~\ref{fig:chivsH}). Above $H_c$, $\beta$-Li$_2$IrO$_3$ shows robust spin-spin correlations that not only give rise to resolution-limited peaks in RXS~\cite{ruiz2017} but also manifest themselves in NMR, which probes the system on a much longer time scale. No inhomogeneities or dynamic spins evading the $Q=0$ correlations are observed.

No clear analogy between the field-induced and pressure-induced~\cite{majumder2018} states of $\beta$-Li$_2$IrO$_3$ can be established. While the former appears upon a second-order phase transition, application of pressure triggers a first-order transformation with phase coexistence around 1.4\,GPa. Magnetic field gradually suppresses the $T_N$, whereas pressure leads to a slight increase in $T_N$ before the ordered state abruptly disappears around 1.4\,GPa. The pressure-induced state is characterized by the absence of local fields~\cite{majumder2018}. On the other hand, local fields develop in the field-induced state below the crossover temperature of about 40\,K. These observations classify the field-induced state as quantum paramagnet, while the pressure-induced state shows signatures of a spin liquid. Similar physics probably occurs in $\gamma$-Li$_2$IrO$_3$, where the incommensurately ordered state can be suppressed by either applied field~\cite{modic2017} or hydrostatic pressure~\cite{breznay2017}.

\acknowledgments
\textit{Acknowledgments.} We thank Manuel Fix for his technical assistance during the pulsed-field measurement and Dana Vieweg for her help with the high-temperature susceptibility measurement. We would also like to thank Hiroshi Yasuoka, Natalia Perkins, and Ioannis Rousochotzakis for fruitful discussions. We also thank ESRF for providing the beamtime at ID22, and acknowledge the support of the HLD at HZDR, member of the European Magnetic Field Laboratory (EMFL). This work was supported by the German Research Foundation (DFG) via projects number 107745057 (TRR80) and JE 748/1 (AJ), and by the Federal Ministry for Education and Research via the Sofja Kovalevskaya Award of Alexander von Humboldt Foundation (AAT).

%

\end{document}